\documentclass{PoS}
\def\Ket#1{||#1 \rangle}
\def\Bra#1{\langle #1||}

\def\endauthors{}
\def\authors#1\endauthors{#1}

\def\be{\begin{equation}}
\def\ee{\end{equation}}
\def\br{\begin{eqnarray}}
\def\er{\end{eqnarray}}
\def\brn{\begin{eqnarray*}}
\def\ern{\end{eqnarray*}}
\def\rf#1{{(\ref{#1})}}
\def\ket#1{|#1 \rangle}
\def\bra#1{\langle #1|}
\def\Ket#1{||#1 \rangle}
\def\Bra#1{\langle #1||}

\def\ie{{\em i.e., }}

\def\lbar{\mbox{$\lambda$\kern-0,450em \vrule width0,35em height1,252ex
depth-1,21ex \kern0,051em}}
\def\dbar{\mbox{d\kern-0,347em \vrule width0,3em height1,252ex depth-1,21ex
\kern0,051em}}
\def\Dbar{\mbox{D\kern-0,735em \vrule width0,3em height0,86ex depth-0,81ex
\kern0,40em}}

\def\up{u_{p}}
\def\vp{v_{p}}
\def\un{u_{n}}
\def\vn{v_{n}}

\def\a {{\alpha}}
\def\b {{\beta}}

\def\w {{\omega}}

\def\O {{{\cal O}}}

\def\ba#1{\begin{array}{#1}}
\def\ea{\end{array}}

\def\be{\begin{equation}}
\def\ee{\end{equation}}
\def\br{\begin{eqnarray}}
\def\er{\end{eqnarray}}
\def\brn{\begin{eqnarray*}}
\def\ern{\end{eqnarray*}}
\def\bit{\begin{itemize}}
\def\eit{\end{itemize}}
\def\bnu{\begin{enumerate}}
\def\enu{\end{enumerate}}

\def\={{\simeq}}

\def\go{\rightarrow  }

\def\rf#1{{(\ref{#1})}}

\def\nn{\nonumber }

\def\ket#1{|#1 \rangle}
\def\bra#1{\langle #1|}
\def\Ket#1{||#1 \rangle}
\def\Bra#1{\langle #1||}
\def\ov#1#2{\langle #1 | #2  \rangle }

\def\2q{{{\{}2{\}}_q}}
\def\3q{{{\{}3{\}}_q}}

\def\betabetago {\stackrel{\mbox{\tiny\,\, $\beta\beta^-$ }}
{\longrightarrow}}

\def\gA{g_{\mbox{\tiny A}}}
\title{${2\nu}$ Double Beta Decay within the Relativistic QRPA }

\ShortTitle{${2\nu}\b\b$-Decay in the Relativistic QRPA}

\author{\speaker{Cl\'audio De Conti}\\
       Campus Experimental de Itapeva, Universidade Estadual Paulista, São Paulo, Brazil\\
       E-mail: \email{conti@itapeva.unesp.br}}

\author{Francisco Krmpoti\'c\\
        Instituto de F\'{\i}sica La Plata, CONICET, Universidad Nacional de La Plata, Argentina\\
        E-mail: \email{krmpotic@fisica.unlp.edu.ar}}

\author{Brett Vern Carlson\\
        Departamento de F\'{\i}sica, Instituto Tecnol\'{o}gico de Aeron\'{a}utica, S\~ao Jos\'e dos Campos, Brazil\\
        E-mail: \email{brett@ita.br}}

\abstract{We perform a self-consistent relativistic QRPA (RQRPA) calculation of
$2\nu\b\b$-decay  based
on  relativistic BCS (RBCS) mean field theory results for
odd-odd intermediate nuclei $^{48}$Sc, $^{76}$As, $^{82}$Br, $^{100}$Tc, $^{128}$I, and $^{130}$I.
The RBCS equations that resemble  the non-relativistic ones are constructed from
a Dirac-Gorkov variational functional.
 We use the parameter set $NL1$ for the $\sigma$, $\omega$ and $\rho$ mesons.
The RQRPA equations  are
solved for the residual $\pi+\rho$ interaction by employing the same parameters used
in the RBCS for the latter meson, and experimental values for the pion and nucleon.
The RQRPA  results  for the ${2\nu}\b\b$ matrix elements
are similar to those obtained within the QRPA and the shell model.}
\FullConference{XXXIV edition of the Brazilian Workshop on Nuclear Physics,\\
        5-10 June 2011\\
        Foz de Igua\c{c}u, Paran\'a state, Brasil}

\begin{document}

\section{Introduction}

In nature there are about 50 nuclear systems in which the single
$\beta$ decay is energetically forbidden, and $\beta\beta$ decay
turns out to be the only possible mode of disintegration. It is the
nuclear pairing force which causes such an "anomaly", by making
the mass of the odd-odd isobar, $(N - 1, Z + 1)$, greater
than the masses of its even-even neighbors, $(N,Z)$ and $(N - 2,Z
- 2)$. The modes by which this decay can take place are connected
with the neutrino ($\nu $)-antineutrino ($\tilde{\nu}$) distinction.
In fact, they
are defined by the transitions:
\begin{eqnarray}
&& n \go p+e^-+{\tilde\nu}_{RH},\\
\nonumber
&& {\nu}_{LH}+n\go p+e^-,
\label{1.1}
\end{eqnarray}
the neutrino ${\nu}$ being left-handed (LH) and the antineutrino $\tilde{\nu}$ right-handed (RH)
  because of parity non-conservation in weak interactions.
Therefore, regardless of the Dirac ($\nu \neq \tilde{\nu}$) or Majorana ($\nu = \tilde{\nu}$)
nature of the neutrino and independently of  conservation of helicity,
the two-neutrino mode ($2\nu\beta\beta$)   decay can occur by two successive single $\beta$-decays:
\begin{eqnarray}
(N,Z) & \stackrel{\beta^-}{\longrightarrow} & (N-1,Z+1) + e^- + \tilde{\nu}\nonumber\\
& \stackrel{\beta^-}{\longrightarrow} & (N-2,Z+2) + 2e^- + 2\tilde{\nu}
\label{1.2}
\end{eqnarray}
passing through the intermediate virtual states of the $(N-1,Z+1)$ nucleus. Yet,
the occurrence of  the neutrinoless $\beta\beta$ decay ($0\nu\beta\beta$):
\begin{eqnarray}
(N,Z) \betabetago (N-2,Z+2) + 2e^-
\label{1.3}
\end{eqnarray}
is much more convoluted since
the right-handed neutrino emitted in the
first step of \rf{1.2} has the wrong helicity to be reabsorbed in a second step and to give rise to
 \rf{1.3}. For massless
neutrinos there is no mixture of left and right handedness, and the $0\nu\beta \beta$-decay
cannot occur, regardless of the Dirac or Majorana nature of the neutrino.
Yet,  experiments with solar, atmospheric and reactor neutrinos  have provided
remarkable evidence in recent years
for the existence of neutrino oscillations driven by nonzero
neutrino masses and neutrino mixing \cite{Gon03,Cam08}.
Once the neutrino becomes massive, the  helicity is no longer a good quantum number.
Then, if the neutrino is in addition a Majorana particle with an effective mass $\langle m_\nu \rangle$,
the mixture of ${\nu}_{LH}$ in ${\nu}_{RH}$ is proportional to $\langle m_\nu \rangle/ E_\nu$,
and  $0\nu\beta \beta$-decay is allowed.
 \footnote{For simplicity we assume that
right-handed weak currents do not play an essential role in the
neutrinoless decay.}
This fact inspired experimental searches in many nuclei,
not only for the ${0\nu}\beta \beta$-decay but also for the ${2\nu}\beta \beta$-decay, since these two modes of
disintegration are related through nuclear structure effects.
In fact, their half-lives can be cast in
the form:
\begin{eqnarray}
T_{2\nu}^{-1}={\cal G}_{2\nu}{\cal M}_{2\nu}^2,
\hspace{2cm}
T_{0\nu}^{-1}={\cal G}_{0\nu}{\cal M}_{0\nu}^2 \langle m_\nu  \rangle^2,
\label{1.4}
\end{eqnarray}
where ${\cal G}'s$ are  geometrical phase space
factors, and  the ${\cal M}'s$ are
nuclear matrix elements (NME's).  ${\cal M}_{2\nu}$ and  ${\cal M}_{0\nu}$
present many similar features, to the extent that it is frequently stated that we
shall not understand
$0\nu\beta \beta$-decay until we understand $2\nu\beta \beta$-decay.

Quantum hadrodynamics (QHD) aims to describe the nuclear many-body
system in terms of nucleons and mesons \cite{Ser86}. Proposed
initially as a fully renormalizable quantum field theory, at present
it is seen as an {\em effective field theory}, derivable, in
principle, from quantum chromodynamics \cite{Ser97}. Relativistic
mean field theory (RMFT), which can be thought as a mean field
(Hartree) approximation to QHD, has been applied with great
success during the last decades to account for nuclear matter
saturation and the ground state properties of finite nuclei along the
whole periodic table. Through a relativistic version of the
random phase approximation (RRPA), various excited states and
the Gamow-Teller
(GT) and Fermi (F) resonances have been studied in this context as well
\cite{Con99,Con00,Vre00,Pie00,Ma01,Ma02,Vre01,Vre01a,Vre02}.
 When this approximation is based on
the Hartree-Bogoliubov (HB) or BCS approximation, it is called the
relativistic quasiparticle RPA (RQRPA). This approach has been used to evaluate several weak
interaction processes, such as beta-decays, neutrino-nucleus reactions,
and muon captures~\cite{Ma07,Pa08,Ma09}. Here the first application
of  the RQRPA to ${2\nu}\b\b$-decay is made.

\section{${2\nu}\beta \beta$ matrix element}

Independently of the nuclear model used and when only allowed transitions are considered,
the ${2\nu}\beta \beta$ matrix element for the $\ket{0^{+}_f}$ final state reads~\cite{Krm05}
\begin{eqnarray}
{\cal M}_{2{\nu}}(f)&=&\sum_{\lambda=0,1}(-)^\lambda
  \sum_{\alpha} \left[\frac
{\Bra{0^{+}_f}\O^{\beta^-}_\lambda\Ket{\lambda^{+}_\alpha}
\Bra{\lambda^{+}_\alpha}\O^{\beta^-}_\lambda\Ket{0^{+}}}
{{E}_{\lambda^{+}_\alpha }-{E_0+E_{0^+_f}}/{2}} \right]
\equiv {\cal M}_{2{\nu}}^{F}(f)+ {\cal M}_{2\nu}^{GT}(f)
\label{2.1}\end{eqnarray}
where $E_0$ and  $E_{0^+_f}$ are, respectively, the
energy  of the initial state $\ket{0^+}$  and
of the final states $\ket{0^+_f}$.
 The summation goes over all intermediate virtual states
$ \ket{ \lambda^{+}_\alpha} $, and
\br
&&\O^{\beta^-}_\lambda=(2\lambda+1)^{-1/2}\sum_{pn}\Bra{p}{\rm O}_\lambda\Ket{n}
\left(c^{{\dagger}}_p c_{\bar{n}}\right)_\lambda,
\hspace{1cm}\mbox{with}\hspace{1cm} \left\{\begin{array}{ll}
{\rm O}_0=1\;\;& \mbox{for F} \;\\
{\rm O}_1=\sigma\;\;& \mbox{for GT} \;\\
\end{array}\right.
\label{2.2}
\end{eqnarray}
are  operators for the $\beta^-$-decay.
The corresponding $\beta^+$-decay operators are
$\O^{\beta^+}_\lambda=\left(\O^{\beta^-}_\lambda\right)^\dagger$. The total $\beta^{\pm}$ strengths
\begin{equation}
 S^{\beta^{\pm}}_\lambda =(2\lambda+1)^{-1}
\sum_{\a}|\Bra{\lambda_\a^+}\O^{\beta^{\pm}}_\lambda\Ket{0^+}|^2,
\label{2.3}
\end{equation}
obey  the well-known single-charge-exchange  Ikeda sum rule (ISR)
\cite{Ike65} for both the F and the GT transitions:
\begin{equation}
S_\lambda^\beta\equiv S^{\beta^{-}}_\lambda-S^{\beta^{+}}_\lambda
=(-)^\lambda
(2\lambda+1)^{-1}\bra{0^+}[\O^{\beta^{+}}_\lambda,
\O^{\beta^{-}}_\lambda]_0\ket{0^+}=N-Z.
\label{2.4}
\end{equation}
Similarly, the  $\beta\beta$-decay strengths
\br
 S^{\beta\beta^{\pm}}_\lambda&=&(2\lambda+1)^{-1}
\sum_f|\bra{0^{+}_f}\O^{\beta^{\pm}}_\lambda\cdot\O^{\beta^{\pm}}_\lambda
\ket{0^{+}}|^2
\label{2.5}\end{eqnarray}
 obey the double-charge-exchange sum rules (DSR):
\br
 S^{\beta\beta}_\lambda =S^{\beta\beta^{-}}_\lambda-S^{\beta\beta^{+}}_\lambda
=(2\lambda+1)^{-1}
\bra{0^{+}}[\O^{\beta^{+}}_\lambda\cdot\O^{\beta^{+}}_\lambda,
\O^{\beta^{-}}_\lambda\cdot\O^{\beta^{-}}_\lambda]\ket{0^{+}},
\label{2.6}\end{eqnarray}
which when evaluated give \cite{Mut92}:
\br
 S^{\beta\beta}_{F}\equiv S^{\beta\beta}_0 &=&2(N-Z)(N-Z-1),
\nn\\
 S^{\beta\beta}_{GT}\equiv  S^{\beta\beta}_1 &=&2(N-Z)\left(N-Z-1+2S^{\beta^+}_1\right)-\frac{2}{3}C,
\label{2.7}\end{eqnarray}
where $C$ is a relatively small quantity and is given by  \cite[Eq. (5)]{Mut92}.
The DSR are as important to $\beta\beta$-decay as
the ISR is for simple $\beta$-decay.

{Contributions from the first-forbidden operators appearing
in the multipole expansion of the weak Hamiltonian, as well as those from
the weak-magnetism term and other second order corrections on the allowed
$2\nu\beta \beta$-decay,  have been examined  rather thoroughly~\cite{Bar95, Bar99}.}
%\section{ Charge-exchange Quasiparticle Random Phase Approximation (QRPA)}

\section{ Charge-exchange QRPA}

The pn-QRPA was formulated and applied to the allowed  $\beta^{\pm}$-decays
 and to the collective GT resonance by Halbleib and Sorensen (HS)  in 1967 \cite{Hal67}.
They solved the  QRPA equation
\begin{eqnarray}
\left(\begin{array}{ll} A & B \\  B &
A\end{array}\right) \left(\begin{array}{l} X
 \\ Y \end{array}\right) =
\omega_\alpha \left(\begin{array}{l} ~X \\-Y \end{array}\right),
\label{3.1} \end{eqnarray}
 within the pn  quasiparticle (qp) space for the  BCS vacuum
\be
\ket{{0}^+}=\prod_p(\up+\vp c^\dag_p c^\dag_{\bar p})
\prod_n(\un+\vn c^\dag_n c^\dag_{\bar n})\ket{},
\label{3.2}\ee
of the initial nucleus
$(N, Z)$, where $\ket{}$ stands for the particle vacuum.
The
 transition matrix elements  are:
\begin{eqnarray}
\Bra{1^{+}_\alpha}\O^{\beta^-}_1\Ket{ 0^{+}}&=&\sum_{pn}
\left[\up \vn X_{{ pn};1^{+}_\alpha}+\vp \un  Y_{pn;1^{+}_\alpha}\right]\Bra{p}{\rm O}_1\Ket{n},
\nn\\
\Bra{1^{+}_\alpha}\O^{\beta^+}_1\Ket{ 0^{+}}
&=&\sum_{pn}
\left[ \vp \un X_{pn;1^{+}_\alpha}+\up \vn Y_{pn;1^{+}_\alpha}\right]\Bra{p}{\rm O}_1\Ket{n},
\label{3.3}
\end{eqnarray}
 the ISR \rf{2.3} yields $N-Z$,
and the ground state correlations (GSC) in \rf{3.3} play an essential
role in  suppressing $\beta^{+}$-decay.

Intensive applications of the
QRPA to $\beta\beta$-decay began only about 20 years later when  Vogel
and Zirnbauer \cite{Vog86} discovered that
 the  $\beta^{+}$-decay suppression mechanism could also be invoked
 to explain the quenching of   the $2\nu\beta \beta$ decay rates. Their adaptation
  of the HS model in essence implies: 1)
To use a second  BCS vacuum for the final nucleus $(N-2, Z+2)$, and to solve a second
QRPA equation for the
intermediate $\overline{1}^+$ states   with the ISR equal to $N-Z-4$,
and 2) To substitute  \rf{2.1}  by the ansatz~\cite{Civ87}:
%\newpage
\begin{eqnarray}
{\cal M}_{2{\nu}}& =&{2} \sum_{\alpha\alpha'}
\frac{\Bra{\overline{1}^+_{\alpha'}}\O^{\beta^+}_1\Ket{\overline{0}^+}
\ov{\overline{1}^+_{\alpha'}}{1^+_\alpha}
\Bra{1^+_\alpha}\O^{\beta^-}_1\Ket{ 0^{+}}} {{\w}_{
1^+_\alpha}+{\w}_{ \overline{1}^+_{\alpha'}}}.
\label{3.4}\end{eqnarray}
%\newpage
To circumvent the nonphysical averaging procedure
implicit in the overlap
$\ov{\overline{1}^+_{\alpha'}}{1^+_\alpha}$, a different  recipe
for  the application of the QRPA to the $\beta \beta$-decay has
been introduced~\cite{Hir90a,Hir90b,Krm90,Krm93,Krm94,Krm94a},
which continues to involve two BCS ground states but deals
with only one set of QRPA solutions.
 Finally, the procedure has been simplified even more by
 performing a straightforward adaptation to the $\b\b$-decay of
   Cha's prescription for the evaluation of single
 $\beta$-decay~\cite{Cha83} within the HS model, which implies solving both the BCS
 and  QRPA equations  for the
intermediate $(N-1, Z+1)$ nucleus.
 The ISR then gives $N-Z-2$, and the above
expression becomes~\cite{Krm97}
\begin{eqnarray}
{\cal M}_{2{\nu}}& =& \sum_{\alpha}
\frac{\Bra{{1}^+_{\alpha}}\O^{\beta^+}_1\Ket{{0}^+}
\Bra{1^+_\alpha}\O^{\beta^-}_1\Ket{ 0^{+}}}
{{\w}_{ 1^+_\alpha}}.
\label{3.5}\end{eqnarray}
Numerical tests show that all the approaches above furnish quite similar results for
${\cal M}_{2{\nu}}$~\cite{Hir90a,Hir90b,Krm93,Krm94,Krm97}. Thus, for the sake of simplicity,
 we make use of the last expression in the present work.

When applied to the $\beta \beta$-decay, the QRPA turns out to be an uncomplete model, since
it deals with 0 and 2 qp states only, while to evaluate Eq.  \rf{1.2}
 it is necessary to consider  at least up to 4 qp states. Moreover, the QRPA can say nothing
 regarding the DSR given by \rf{2.7}, nor it can be used to describe the
$\beta \beta$-decays to excited final states.
 How this can be implement is explained in Ref.~\cite{Krm05}.

\section{Relativistic charge-exchange RQRPA}

The RQRPA is based on the relativistic HB (RHB) approximation for the RMFT.
 It was formulated in Ref.~\cite{Pa03} for charge-conserving excitations,
and extended
to  charge-exchange excitations in Ref.~\cite{Pa04}.
In the present work we approximate the RHB
equations and put them in a form that resembles the
non-relativistic BCS equations. To do this we start from the variational functional
\begin{eqnarray}
W & = &\int d^{3}xd^{3}y\sum_{t}(U_{t}^{\dagger }({\bold
x}),V_{t}^{\dagger }({\bold x})\gamma _{0})
\nonumber\\
& \times &\left(
\begin{array}{cc}
(\omega_{t}+\mu _{t})\delta ({\bold x}-{\bold y})-h_{t}({\bold
x},{\bold y}) & \overline{\Delta }_{t}^{\dagger }({\bold x},{\bold y})
\\
\overline{\Delta }_{t}({\bold x},{\bold y}) & (\omega _{t}-\mu _{t})\delta ({\bold x}-%
{\bold y})-h_{t}({\bold x},{\bold y})
\end{array}
\right)
\left(
\begin{array}{c}
U_{t}({\bold x}) \\
\gamma _{0}V_{t}({\bold x})
\end{array}
\right), \label{4.1}
\end{eqnarray}
based on the Dirac-Gorkov equation \cite[Eq.(39)]{Ca00} with notation $t=p$ or $n$.
Here $U_{t}({\bold x})$ and $V_{t}({\bold x})$ are the normal
and time-reversed Dirac spinors corresponding to  solutions of this
equation with  positive and
negative-frequency $\omega _{t}$. The Lagrange multipliers $\mu_t$ are  determined by
requiring that the expectation values of the baryon number
operators yield the desired values of $Z$
and $N$.
Dirac Hamiltonian operators $h_{t}({\bold x},{\bold y})$
and pairing fields
$\overline{\Delta }_{t}({\bold x},{\bold y})$ are given by
\cite[Eqs. (40) and (49)]{Ca00}. Next we  use  the anzatz
\begin{eqnarray}
\left(
\begin{array}{c}
U_{t}({\bold x}) \\
\gamma _{0}V_{t}({\bold x})
\end{array}
\right) \longrightarrow \left(
\begin{array}{c}
u_{t}{\cal U}_{t}({\bold x}) \\
v_{t}{\cal U}_{t}({\bold x})
\end{array}
\right), \label{4.2}
\end{eqnarray}
where $u_{t}$ and $v_{t}$ are numbers, and ${\cal U}_{t}({\bold
x})$ are the  Hartree mean-field  wave functions, satisfying
the equation $\int d^3y h_t ({\bold x},{\bold y}){\cal U}_{t}({\bold y}) =
\varepsilon_{t} {\cal U}_{t}({\bold x})$, with
 $\varepsilon_{t}$ being the single-particle energies.
After performing this replacement
 one  maximizes (\ref{4.1})  with respect
to the coefficients $u_{t}$, obtaining in this way the relativistic
BCS (RBCS) equations for $u_{t}$ and $v_{t}$, similar to  the non-relativistic ones.

Since the pion does not participate in the RBCS,  the Lagrangian density
is determined once the masses of the nucleon and mesons $\sigma$, $\omega$ and $\rho$,
the coupling constants of mesons with the nucleon, and the self-interaction
constants of the meson $\sigma$, g2 and g3 are given.
Several sets for these parameters are known in the literature. Here we use the
 $NL1$~ set\cite{Rei}.

We solve the  RBCS and the Klein-Gordon equations  numerically by expanding the
mesons fields and the fermions wave functions in complete sets of eigenfunctions
of harmonic oscillator (HO) potentials. In actual calculations,
the expansion is truncated at a finite number of major shells, with the
quantum number of the last included shell denoted by $N_F$ ($N_B$) for fermions
 (bosons). The maximum values are selected so as to assure the
physical significance of the results.
The oscillator frequency for fermions is given by $\hbar\omega_0 = 41A^{-1/3}$ MeV
and the maximum number of oscillator shells for fermions and bosons is given
by $N_F=N_B=20$. The Coulomb field is calculated directly in configuration space.
The same procedure was used by Ghambir {\it et al}~\cite{Gam}
 in their approach to the relativistic mean field.

The RQRPA equations (\ref{3.1}) are solved by employing for the residual interaction
the same
 parameters used in the RMFT to obtain the discrete basis of qp states
within the RBCS approximation. Yet, in dealing with isovector excitations it is essential to
include, together with the $\rho$ meson, the $\pi$ meson as well~\cite{Con99,Con00}.
Here, the experimental values of the pseudoscalar pion-nucleon coupling and the
 pion mass  were used, \ie  $f_\pi=1.00$, and $m_\pi =138.0$ MeV~\cite{Con99,Con00}.
 Since Fock terms are ignored
 in the RMFT, for the sake of self-consistency  we must omit the exchange matrix
 element of the residual interaction $V=V_\pi+V_\rho$ in the sub-matrices $A$ and $B$.

\section{Results}
%%%%%%%%%%%%%%
\begin{table}[h]
\caption{Matrix elements ${\cal M}_{2\nu}$ in units of (MeV)$^{-1}$. The experimental
values are obtained from the half-lives $T_{2\nu}$ compiled by Barabash~\cite{Bar10}
and the ${\cal G}_{2\nu}$ values from Ref.~\cite{Doi85} for the bare axial-vector coupling constant
$\gA^0=1.25$.}
\begin{minipage}[h]{12cm}
\begin{tabular}{cccccc} \hline \hline\
$\b\b$-Decay &EXP &SM~\cite{Cau11}&SMren~\cite{Cau11}&QRPA~\cite{Krm94}&RQRPA \\
\hline
$^{48}$Ca $\go^{48}$Sc $\go^{48}$Ti      &$   0.049\pm   0.003$&$   0.047$&$   0.055$&$   0.058$&$  0.066  $\\
$^{76}$Ge $\go^{76}$As $\go^{76}$Se       &$   0.140\pm   0.005$&$   0.116$&$   0.206$&$   0.064$&$ 0.055  $\\
$^{82}$Se $\go^{82}$Br $\go^{82}$Kr      &$   0.098\pm   0.004$&$   0.126$&$   0.224$&$   0.077$&$  0.088  $\\
$^{100}$Mo$\go^{100}$Tc$\go^{100}$Ru      &$   0.239\pm   0.007$&$   -$&$  -$&$   0.065$&$  0.062 $ \\
$^{128}$Te$\go^{128}$I$\go^{128}$Xe      &$   0.049\pm   0.006$&$   0.059$&$   0.116$&$   0.076$&$  0.076  $\\
$^{130}$Te$\go^{130}$I$\go^{130}$Xe     &$   0.034\pm   0.003$&$   0.043$&$   0.085$&$   0.061$&$   0.070  $\\
\hline\hline
   \end{tabular}
\label{tab1}
\end{minipage}
\end{table}

In Table \ref{tab1} the experimental values of the matrix elements ${\cal M}_{2\nu}$ are
compared with values obtained from several calculations.
The later critically depend on the adopted value for the
effective axial-vector coupling constant $\gA$.
The QRPA results~\cite{Krm94}, as well as the present ones correspond to
$\gA=1$, \ie to a quenching factor of $q=\gA/\gA^0=0.8$ when the bare value
is $\gA^0=1.25$~\cite{Doi85}.
Bearing in mind that only  a very tiny
fraction ($<0.1\%$) of the sum rule  $S^{\beta\beta}_{GT}$ goes into the final
state $J^\pi_f=1^+_1$~\cite{Krm05}, we can say that both
agree reasonably well with experiment.
One should also remembered that the QRPA calculations
have been monitored by the restoration of the $SU(4)$ symmetry while the RQRPA were not.
In the shell model (SM) study~\cite{Cau11} different $q$ values were used in
different nuclei, namely $q=0.74$ in $^{48}$Ca, $q=0.60$ in  $^{76}$Ge, and $^{82}$Se,
and $q=0.57$ in $^{128}$Te, and $^{130}$Te. These results are listed in the third column of
Table \ref{tab1}. For the sake of comparison,  the same results renormalized to $q=0.8$
are shown in the fourth column (labelled as  "SMren").

\section{Conclusion}
%The RQRPA is build up on the

A variational functional based on the Dirac-Gorkov equation is used to obtain the RHB
equations in the form of the non-relativistic BCS equations.
The RQRPA equations (\ref{3.1}) are
solved for the residual $\pi+\rho$ interaction by employing for the latter meson
the same parameters used in the RMFT.
The RQRPA  results  for the ${2\nu}\b\b$ matrix elements
are of the same order magnitude as those obtained within the QRPA and the SM.
 Bearing in mind the small fraction of double GT strength
 going to the $0^+$ final state compared with the GT DSR, as well as the uncertainty involved
 in the quenching factor q, it is difficult to discern which of the three calculations
 is better and which is worse.
Despite this we are planning to  apply  the RQRPA model to study the neutrinoless
 $\b\b$-decays using the formalism developed in Ref.~\cite{Bar99} as well.
\section*{Acknowledgements}
This work was partially supported by the Argentinean agency CONICET under
contract PIP 0377. BVC acknowledeges partial support from FAPESP and the CNPq.


\begin{thebibliography}{99}
\bibitem{Gon03}  M. C. Gonzalez-Garcia and  Y. Nir, \emph{Rev. Mod. Phys.} {\bf 75} (2003) 345 .
\bibitem{Cam08} L. Camilleri, E. Lisi and J. F. Wilkerson, \emph{Ann. Rev. Nucl. Part. Sci.} {\bf 58} (2008) 343.
\bibitem{Ser86}{B.  D.  Serot and J.  D.  Walecka, \emph{Adv.  Nucl.  Phys.} {\bf 16} (1986) 1.}
\bibitem{Ser97}{B.  D.  Serot and J.  D.  Walecka, \emph{Int.  J.  Mod. Phys.} {\bf E 6} (1997) 515.}
\bibitem{Con99}{C. De Conti, A. P. Gale\~ao and F. Krmpoti\'c, \emph{Phys. Lett.} {\bf B 444} (1999) 14.}
\bibitem{Con00}{C. De Conti, A. P. Gale\~ao and F. Krmpoti\'c, \emph{Phys. Lett.} {\bf B 494} (2000) 46.}
\bibitem{Vre00}{D. Vretenar, A. Wandelt and P. Ring, \emph{Phys. Lett.} {\bf B 487} (2000) 334.}
\bibitem{Pie00}{J. Piekarewicz, \emph{Phys.Rev.} {\bf C 64} (2001) 024307.}
\bibitem{Ma01}{Z. Y. Ma, N. Van Giai, A. Wandelt, D. Vretenar and P. Ring, \emph{Nucl. Phys.} {\bf A 686} (2001) 173.}
\bibitem{Ma02}{Z. Y. Ma,  A. Wandelt,  N. Van Giai,  D. Vretenar, P. Ring and L.G. Cao, \emph{Nucl. Phys.} {\bf A 703} (2002) 222.}
\bibitem{Vre01}{D. Vretenar, N. Paar, P. Ring, and G. A. Lalazissis, \emph{Phys. Rev.} {\bf C 63} (2001) 047301.}
\bibitem{Vre01a}{D. Vretenar, N. Paar, P. Ring, and G. A. Lalazissis, \emph{Nucl. Phys.} {\bf A 692} (2001) 496.}
\bibitem{Vre02}{D. Vretenar, N. Paar, T. Nik\v{s}i\'c, and P. Ring, \emph{Phys. Rev.} {\bf C 65} (2002) 021301.}
\bibitem{Ma07}{T. Marketin, D. Vretenar and P. Ring, \emph{Phys. Rev.} {\bf C 75} (2007) 024304.}
\bibitem{Pa08}{N. Paar, D. Vretenar, T. Marketin, and P. Ring, \emph{Phys. Rev.} {\bf C 77} (2008) 024608.}
\bibitem{Ma09}{T. Marketin, N. Paar,T. Nik\v{s}i\'c and D. Vretenar, \emph{Phys. Rev.} {\bf C 79} (2009) 054323.}
\bibitem{Krm05} {F. Krmpoti\'c, \emph{FIZIKA} {\bf B 14} (2005) 2, 139.}
\bibitem{Ike65} K. Ikeda, T. Udagawa, H. Yamaura, \emph{Prog. Theor. Phys.} {\bf 175} (1965) 22.
\bibitem{Mut92} K. Muto, \emph{Phys. Lett.} {\bf B 277} (1992) 13.
\bibitem{Bar95} C. Barbero, F. Krmpoti\'c and  A. Mariano, \emph{Phys. Lett.} {\bf B 345} (1995) 192, {\it ibid} \emph{Phys. Lett.} {\bf B 436} (1998) 49.
\bibitem{Bar99} C. Barbero, F. Krmpoti\'c, A. Mariano and  D. Tadi\'c, \emph{Nucl. Phys.} {\bf A 650} (1999) 485.
\bibitem{Hal67} J. A. Halbleib and  R. A. Sorensen, \emph{Nucl. Phys.} {\bf A 98} (1967) 524.
\bibitem{Vog86} P. Vogel and  M. R. Zirnbauer, \emph{Phys. Rev. Lett.} {\bf 57} (1986) 731.
\bibitem{Civ87} O. Civitarese, A. Faessler and  T. Tomoda, \emph{Phys. Lett.} {\bf B 194} (1987) 11.
\bibitem{Hir90a} J. Hirsch and  F. Krmpoti\'c, \emph{ Phys. Lett. } {\bf B 246} (1990) 5.
\bibitem{Hir90b} J. Hirsch and F. Krmpoti\'{c}, \emph{Phys. Rev.} {\bf C 41} (1990) 792.
\bibitem{Krm90} F. Krmpoti\'{c}, in Lectures on Hadron Physics, Ed. by E. Ferreira
 (World Scientific, Singapore, 1990) p. 205
\bibitem{Krm93} F. Krmpoti\'{c}, A. Mariano, T. T. S. Kuo and  K. Nakayama, \emph{Phys. Lett.} {\bf B 319} (1993) 393.
\bibitem{Krm94} F. Krmpoti\'{c} and S. Shelly Sharma, \emph{Nucl. Phys.} {\bf A 572} (1994) 329.
\bibitem{Krm94a} {F. Krmpoti\'c, \emph{Rev. Mex. F\'{\i}s.} {\bf 40} (1994) 285.}
 \bibitem{Cha83} D. Cha, \emph{Phys. Rev.} {\bf C 27} (1983) 2269.
 \bibitem{Krm97}{F. Krmpoti\'c, T. T. S. Kuo, A. Mariano, E. J. V. de Passos and  A. F. R. de Toledo Piza, \emph{Nucl. Phys.} {\bf A 612} (1997) 223.}
%\bibitem{Ej0} {H. Ejiri, \emph{Phys. Rep.} {\bf 388} (2000) 412.}
\bibitem{Ca00}{B. V. Carlson and D. Hirata, \emph{Phys. Rev.} {\bf C 62} (2000) 054310. }
\bibitem{Pa03}{N. Paar, P. Ring, T . Nik\v{s}i\'c, and D. Vretenar, \emph{Phys. Rev.} {\bf C 67} (2003) 034312.}
\bibitem{Pa04}{N. Paar, T. Nik\v{s}i\'c, D. Vretenar, and P. Ring, \emph{Phys. Rev.} {\bf C 69} (2004) 054303 .}
\bibitem{Rei} {P. G. Reinhard, M. Rufa, J. Maruhn, W. Greiner and J. Friedrich, \emph{Z. Phys.} {\bf A 323} (1986) 13.}
\bibitem{Gam} {Y. K. Gambhir, P. Ring and A. Thimet, \emph{Ann. Phys. (N.Y.)} {\bf 198} (1990) 132.}
\bibitem{Bar10}{A. S. Barabash, \emph{ Phys. Rev.} {\bf C 81} (2010) 035501. }
\bibitem{Doi85}{ M. Doi, T. Kotani and E. Takasugi,
 \emph{Prog. Theor. Phys. Suppl.}{\bf C 83} (1985) 1.}
\bibitem{Cau11} E. Caurier, F. Nowacki, and A. Poves, arXiv:1112.5039.
\end{thebibliography}
\end{document}